\documentclass[twocolumn,prl,superscriptaddress,floatfix,showpacs]{revtex4-1}
\usepackage[utf8]{inputenc}
\usepackage{amsmath}
\usepackage{amssymb}
\usepackage{graphicx}

\makeatletter
\usepackage{graphics}
\usepackage{epsfig}
\usepackage{color}

\usepackage[normalem]{ulem}
\newcommand{\be}{\begin{equation}} \newcommand{\ee}{\end{equation}}
\newcommand{\bea}{\begin{eqnarray}} \newcommand{\eea}{\end{eqnarray}}

\begin{document}
\title{Revisiting a Low-Dimensional Model with Short Range Interactions and Mean Field Critical Behavior}
\author{Peter Grassberger} \affiliation{JSC, FZ J\"ulich, D-52425 J\"ulich, Germany}
\date{\today}

\begin{abstract}
In all local low-dimensional models, scaling at critical points deviates
from mean field behavior -- with one possible exception. This exceptional
model with ``ordinary" behavior is an inherently non-equilibrium model studied 
some time ago by H.-M. Br\"oker and myself. In simulations, its 2-dimensional 
version suggested that two critical exponents were mean-field, while a 
third one showed very small deviations. Moreover, the numerics agreed almost 
perfectly with an explicit mean field model. In the present paper we 
present simulations with much higher statistics, both for 2d and 3d. In both
cases we find that the deviations of all critical exponents from their mean field 
values are non-leading corrections, and that the scaling is {\it precisely} of 
mean field type. As in the original paper, we propose that the mechanism for this 
is ``confusion", a strong randomization of the phases of feed-backs that can occur 
in non-equilibrium systems.
\end{abstract}
\maketitle

Non-linear low-dimensional stochastic systems with short range interactions tend
to show anomalously large fluctuations. These fluctuations then lead to ``anomalous"
scaling laws which deviate from their mean field behavior. This applies to virtually 
all sorts of systems: To rough surfaces (e.g. Kardar-Parisi-Zhang \cite{KPZ} and 
quenched Edwards-Wilkinson \cite{qEW} models), to self-organized critical models 
like the Bak-Tang-Wiesenfeld \cite{BTW} and Manna \cite{Manna} sand pile models and 
the Bak-Sneppen evolution model \cite{BS}, to heat conduction in 1-d systems such 
as the Fermi-Pasta Ulam \cite{FPU} and alternating mass hard particle \cite{GNY}
systems, and -- maybe most famously -- to second order phase transitions as e.g.
in Ising, XY, or Heisenberg models \cite{Amit}, in percolation \cite{Stauffer}, 
and in self-avoiding walks \cite{de_Gennes}.

The possibility of ``normal", i.e. mean field type behavior has been much 
discussed in 1-d heat conduction, where ``normal" behavior would correspond to
the validity of Fourier's law \cite{Casati, Garrido,Gendelman_2000,Gendelman_2014,Yang-G,Casati-Li,Lepri},
but not in any of the other types of phenomena. It is true that in some cases 
even deviations from power law scaling have been observed (as in the Drossel-Schwabl
forest fire model \cite{Pruessner,Grass_DS}), but normal (i.e. mean-field)
scaling at critical points were {\it never} reported -- with one single 
exception. This exception is an old paper by H.-M. Br\"oker and myself 
\cite{Broker}. This paper was cited only 5 times within 26 years (according to 
Google Scolar), which illustrates most clearly that the concept of a low-dimensional
local model with mean field type scaling at a critical point was considered as 
completely outlandish by the community.

The model studied in \cite{Broker} was a modification of Manna's sandpile 
model (for a similar model, see \cite{Manna-KK}), with enhanced stochasticity 
and with non-conservation of `sand'. The latter is controlled by an explicit
control parameter, which changes it from being self-organized critical into
a model with a standard non-equilibrium second-order (continuum) phase transition.
As in most such models, its critical behavior is characterized by three independent
exponents. Two of these were found in \cite{Broker} to be mean field like,
while the third showed very small deviations. Moreover, not only the exponents 
but also the scaling functions were extremely close to those of an explicit
mean field model, namely to a version of the model on a Bethe tree. We should 
add that this model should, by standard arguments \cite{Manna-KK}), be in the 
universality class of the fixed-energy Manna sand pile.

The closeness to mean field suggested of course that the deviations from it could
be finite-size corrections, but simulations at that time were unable to decide
this question. We thus decided to revisit the problem and to perform much larger
simulations with modern hardware. The results presented below are quite unambiguous: 
It seems that all deviations are indeed due to finite-size corrections, both for 
the 2-dimensional version of the model studied in \cite{Broker} and for its 
generalization to 3 dimensions.

This model is defined on a $d$--dimensional hypercubic lattice (generalizations
to other types of lattices are obvious), and time is discrete. At each lattice
site we have a ``spin" $z_{i}$, which can take any non-negative integer
value, but only the values $z_i=0$ and $z_i=1$ are "stable". If $z_i$
becomes $>1$ during the evolution, it ``topples".
The toppling rule is
\be
    z_{i}\; \to \; z_{i}-2                       \label{tpe}
\ee
for the site which topples, and
\be
   z_{j} \to \left\{ \begin{array}{l@{\quad:\quad}l}
          z_{j}+1 & \mbox{with probability}\quad p\\
          z_{j}   & \mbox{with probability}\quad 1-p
                  \end{array}    \right.           \label{tpz}
\ee
for each of its $2d$ neighbors, with $0\leq p \leq 1/d$. Notice that each 
neighbor $j$ has the same chance $p$ to get its spin increased, {\it independently} 
of what happens at the other neighbors. Thus the sum $\sum_iz_i$
fluctuates (during each toppling, it can change by any value between
$-2$ and $2d-2$), but in the average each toppling event causes
$\sum_iz_i$ to decrease by $2dp-2$. The critical
point is exactly where this vanishes, $p_c = 1/d$. For later use we
define $\epsilon=1/d-p$, and $\varrho = \langle z \rangle$ \cite{footnote1}.

As in the sandpile model, the dynamics actually consists of the following rules: \\
(i) We start with a configuration where all sites are stable.\\
(ii) A site $i$ is chosen randomly, and $z_i$ is increased by one 
unit. In the following, we call this an ``event".\\
(iii) If at least one site is unstable, the above toppling rule is applied
immediately (i.e. without increasing the time counter) and simultaneously 
at all unstable sites \cite{footnote2}. After this, $t$ is increased by 1 unit.
If some $z$'s are still $\geq 4$ so that they have to topple again, then 
all these topplings are also done simultaneously. After each such round of 
topplings, $t$ is again increased by 1. This is repeated
until all sites are stable again, after which rule (ii) is applied again.

As we have already said, this is very similar to the model of \cite{Manna-KK}.
The main difference is that we increase the neighboring `spins' independently
during a toppling, which adds to the randomness of the process. We believe 
that it is this enhanced degree of stochasticity which is responsible for 
the very special features of the model.

Although we could also run the model for $p\geq 1/2d$, if we would use open
boundary conditions (as in the original versions of sandpile models), we show
here only results for periodic (more precisely, helical) boundary conditions.
In all simulations, sufficiently long transients were discarded so that all
measurements are for the statistically stationary state. The number of events
were for each pair of values $(d,\epsilon)$ more than $6\times 10^8$.

\begin{figure}
\begin{center}
   \vglue -3.2cm
   \includegraphics[width=0.51\textwidth]{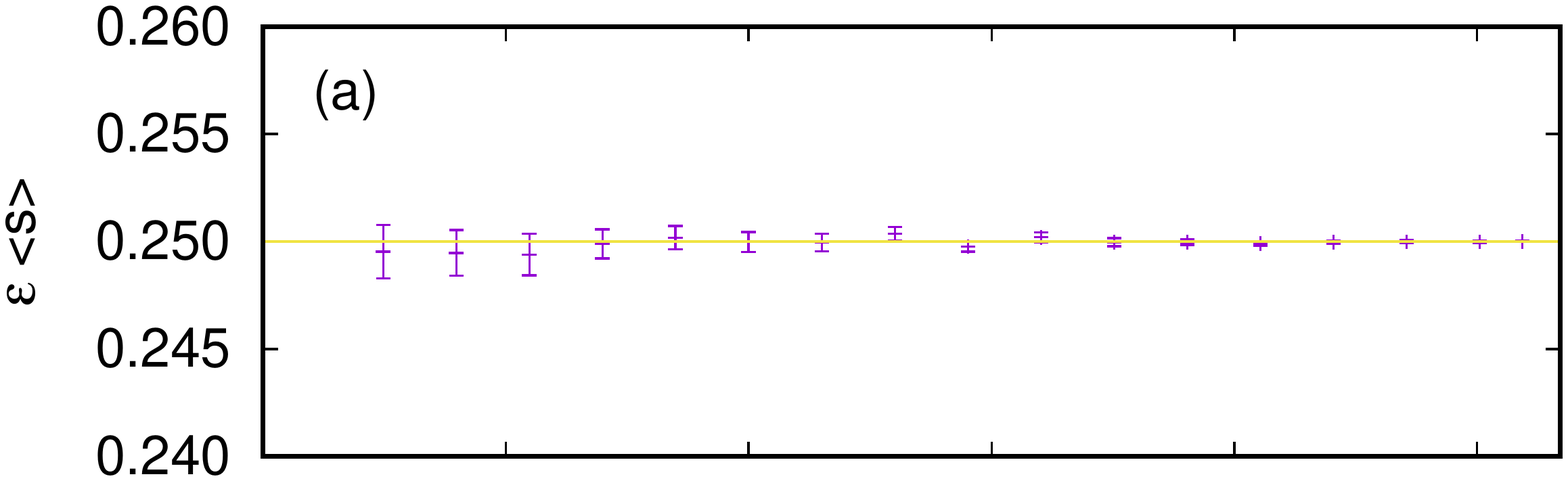}
   \vglue -3.9cm
   \includegraphics[width=0.51\textwidth]{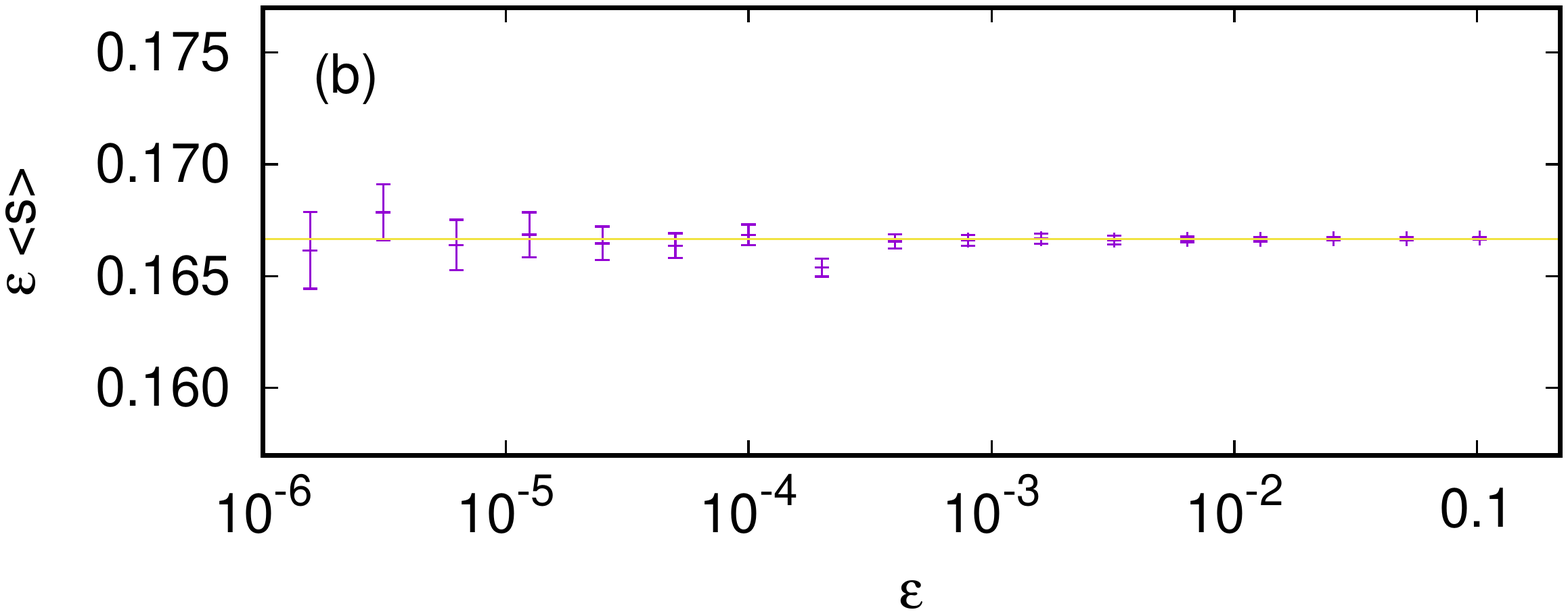}
\end{center}
   \vglue -0.9cm
	\caption{\small Average avalanche sizes (number of topplings per 
	event) plotted against $\epsilon$, for (a) $d=2$ (top panel) and (b) $d=3$
	(bottom panel). The horizontal lines indicate the theoretical predictions.}
        \label{sizes.fig}
\end{figure}

\begin{figure}
\begin{center}
   \vglue -.9cm
   \includegraphics[width=0.51\textwidth]{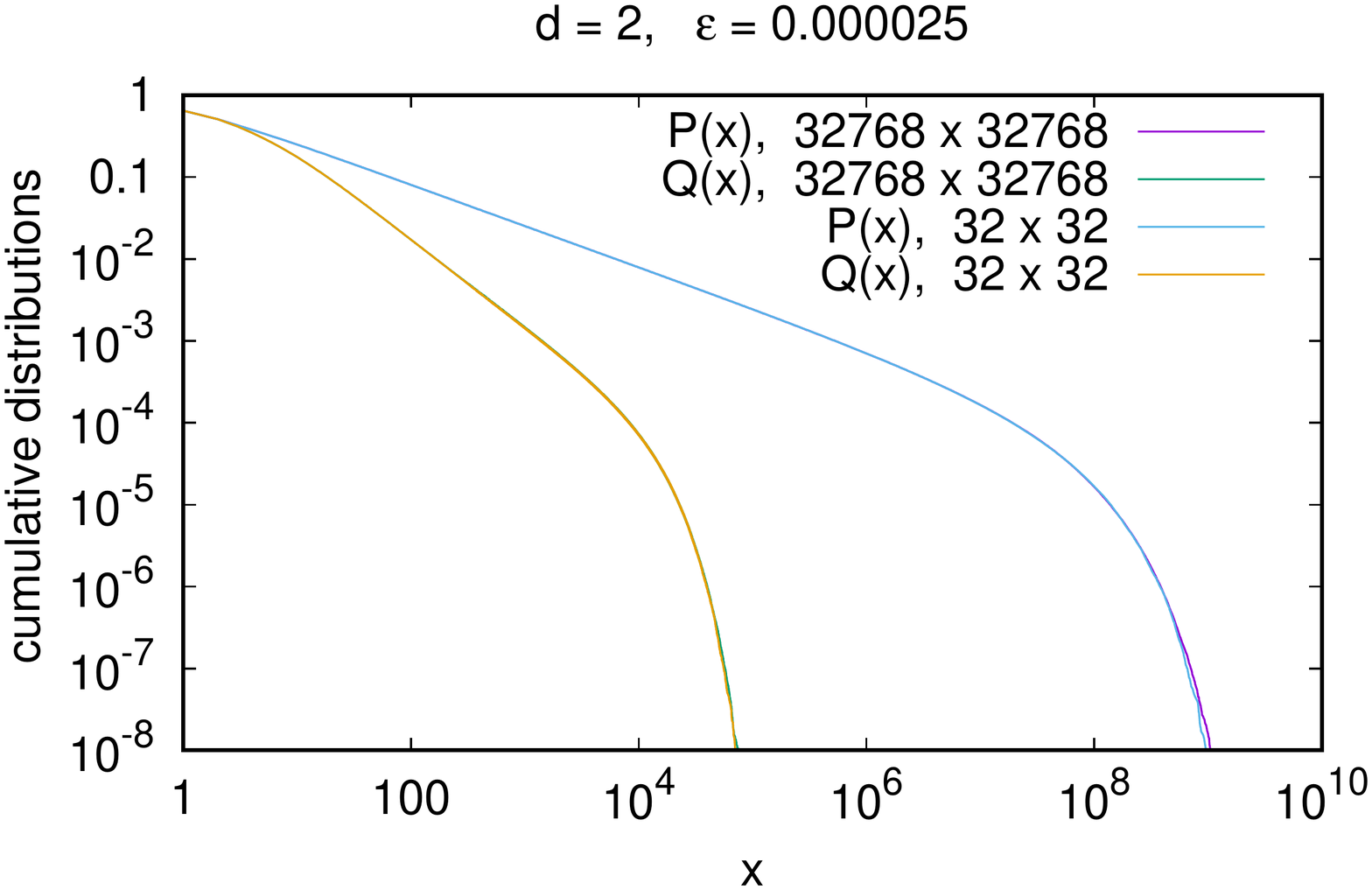}
\end{center}
   \vglue -0.9cm
        \caption{\small Cumulative distributions of the number of topplings per event,
	($P(x)$, upper curve) and of their durations ($Q(x)$, lower curve), for $d=2$ 
	and $\epsilon=0.000025$. Even though only two curves seem to be visible, four 
	curves are actually shown: Each distribution is shown for $L=32$ and for $L=32768$,
	where the simulation boxes have sizes $L\times L$.}
        \label{ifinitesize.fig}
\end{figure}

Since $\sum_i z_i$ decreases in average by $2d\epsilon$ during each toppling, 
while it first is increased by 1 in each event, the average number of topplings 
per event is just $1/(2d\epsilon)$. Verifying this (see Fig.~1) presents thus
a stringent test that stationarity had been reached. 

\begin{figure}
\begin{center}
   \vglue -1.8cm
   \includegraphics[width=0.50\textwidth]{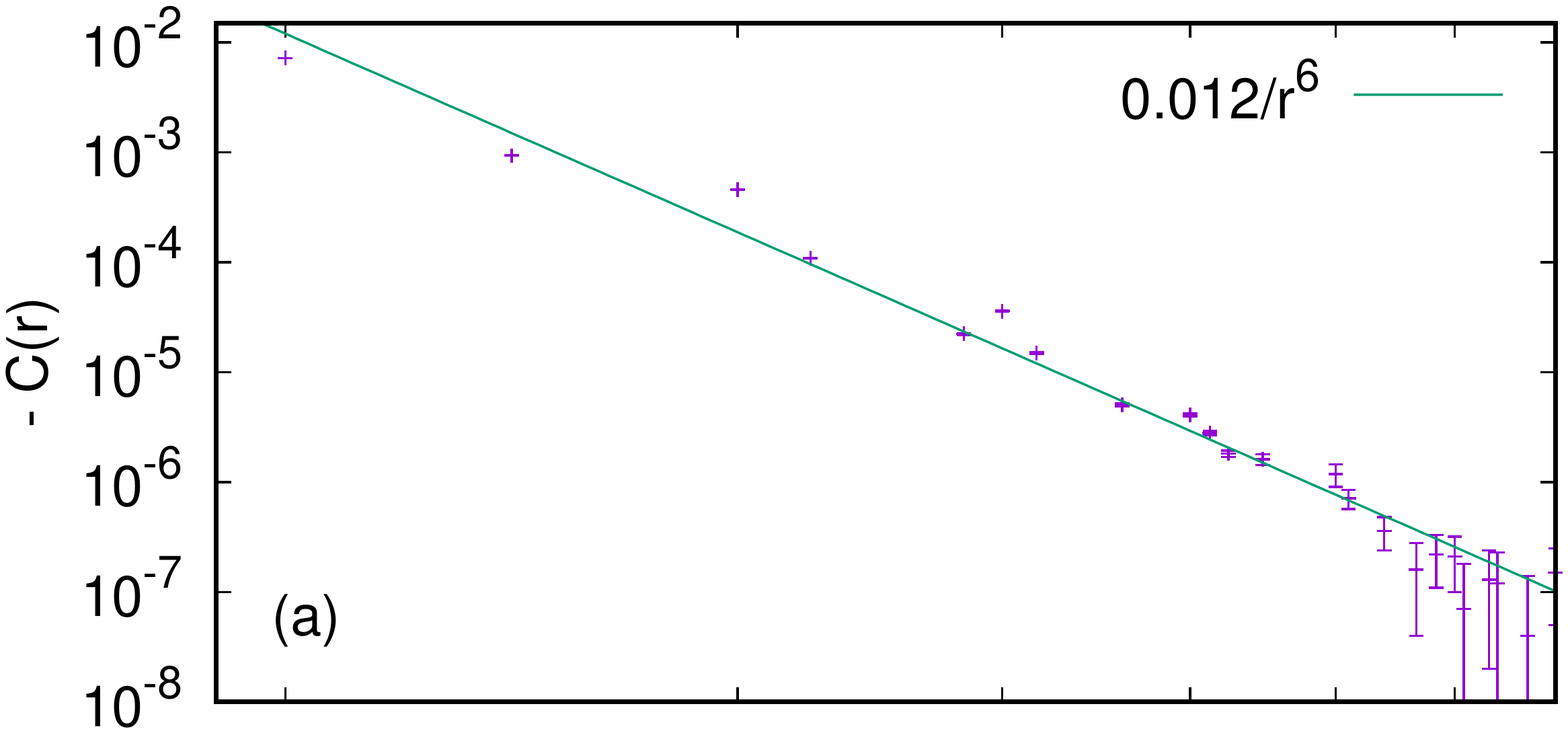}
   \vglue -2.8cm
   \includegraphics[width=0.50\textwidth]{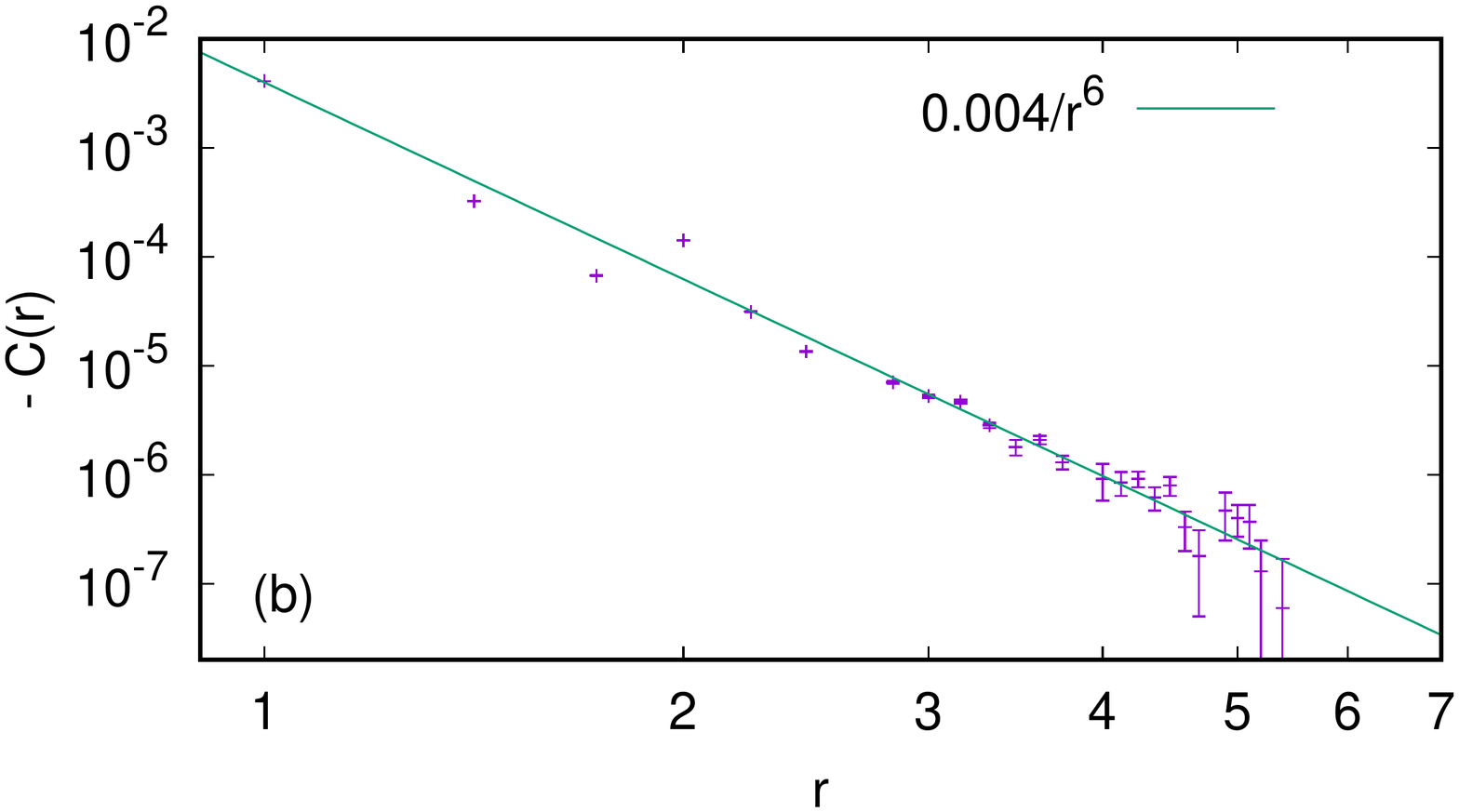}
\end{center}
   \vglue -0.9cm
	\caption{\small Spatial correlations between `spin' values (or densities,
	in a sandpile terminology) distances $r$ apart. Since all correlations for
	all $r>0$ are negative, we show a log-log plot for $-C(r)$ vs. $r$. The straight
	lines give rough power law fits, but notice that deviations from these power
	laws are significant and rather erratic. As in Fig.~1 and in the following 
	figures, panel (a) is for $d=2$, while (b) is for $d=3$.}
        \label{correl.fig}
\end{figure}

This does not, however, test against finite size corrections. But no such 
corrections whatsoever were seen, if we changed the simulation volume between 
$32\times 32$ and $32768\times 32768$ in 2 dimensions, and between $16^3$ and 
$512^3$ in $d=3$ (see Fig.~2). This complete absence of visible finite size effects 
is very surprising (for many sandpile models, finite size effects are huge, 
see e.g. \cite{qEW}). It allowed us to use rather modest lattice sizes: In $d=2$,
most simulations were done with $L = 256$ or 512, and in $d=3$ we used mostly
$L = 128$ and 256 -- although the largest simulated avalanches were huge in both 
cases, and had $\approx 10^{11}$ topplings. 
It is easily explained by the smallness and short ranges of correlations. As 
seen from Fig.~3, these correlations are negative both in $d=2$ and $d=3$, and 
\be
   c(i-j)\equiv\langle z_iz_j\rangle-\varrho^2 \sim ||i-j||^{-6}
\ee
in both dimensions, which is an even faster decay than in the Bak-Tang-Wiesenfeld 
sandpile model (where $c(i-j) \sim ||i-j||^{-4}$ \cite{dhar2}). Notice that the 
absolute values of the correlations are much smaller in $d=3$ than in $d=2$.
This is expected if the behavior is close to mean field, because deviations 
from it should decrease with $d$.
  
Average stationary densities $\varrho$ versus $\epsilon$ are shown in Fig.~4. We
also show there predictions from a mean field model, following \cite{Broker}. 
In this model, a site `remembers' for the present time step that it had toppled, but 
`forgets' it thereafter. Thus a site that had not toppled during the present 
avalanche has density $\varrho$. A toppling site which is not at the root of an 
avalanche has thus $2d-1$ neighbors with density $\varrho$, while its `father' (the
site which made it topple) has a different density $\varrho'$. In the simplest version, 
we also neglect the possibility that the father might have changed after this toppling,
which implies $\varrho'=0$, in which case the problem is essentially that of percolation
on a Bethe lattice with coordination number $2d$ \cite{Stauffer}. But we can also 
allow values $\varrho'\neq 0$, in which case the model can still be solved exactly 
\cite{Broker}. In any case we expect $\varrho'\ll \varrho$. 
Let us denote by $a=p\varrho$ and $a'=p\varrho'$ the probabilities that a toppling (non-root)
site will make its neighbors topple, while the very first toppling of an avalanche
will make all its neighbors topple with probability $a$. The average number 
of topplings during an event (whether it triggered an avalanche or not) is \cite{Broker}
\be
   \langle s(p)\rangle=\varrho[1+2da\sum_{i=0}^{\infty}
     (2da+a'-a)^i] = \varrho \frac{1+a-a'}{1-(2d-1)a-a'}\,.        \label{sp}
\ee
Since we know that $\langle s(p)\rangle = 1/(2d\epsilon)$, we obtain
\be
    \varrho=\frac{1}{2}+\frac{2a(a-a')}{1+(2d-1)a-a'},\quad \epsilon = 1/d-a/\varrho.
\ee
This allows us to obtain $\varrho$ as a function of $\epsilon$ for any fixed ratio $a'/a$.
From Fig.~4 we see that very good fits are obtained in both dimensions with $a'=0$. 
While this is indeed the best fit in $d=3$, an even better fit is obtained in $d=2$ with 
$a'\sim a^3$.

\begin{figure}
\begin{center}
   \vglue -1.9cm
   \includegraphics[width=0.50\textwidth]{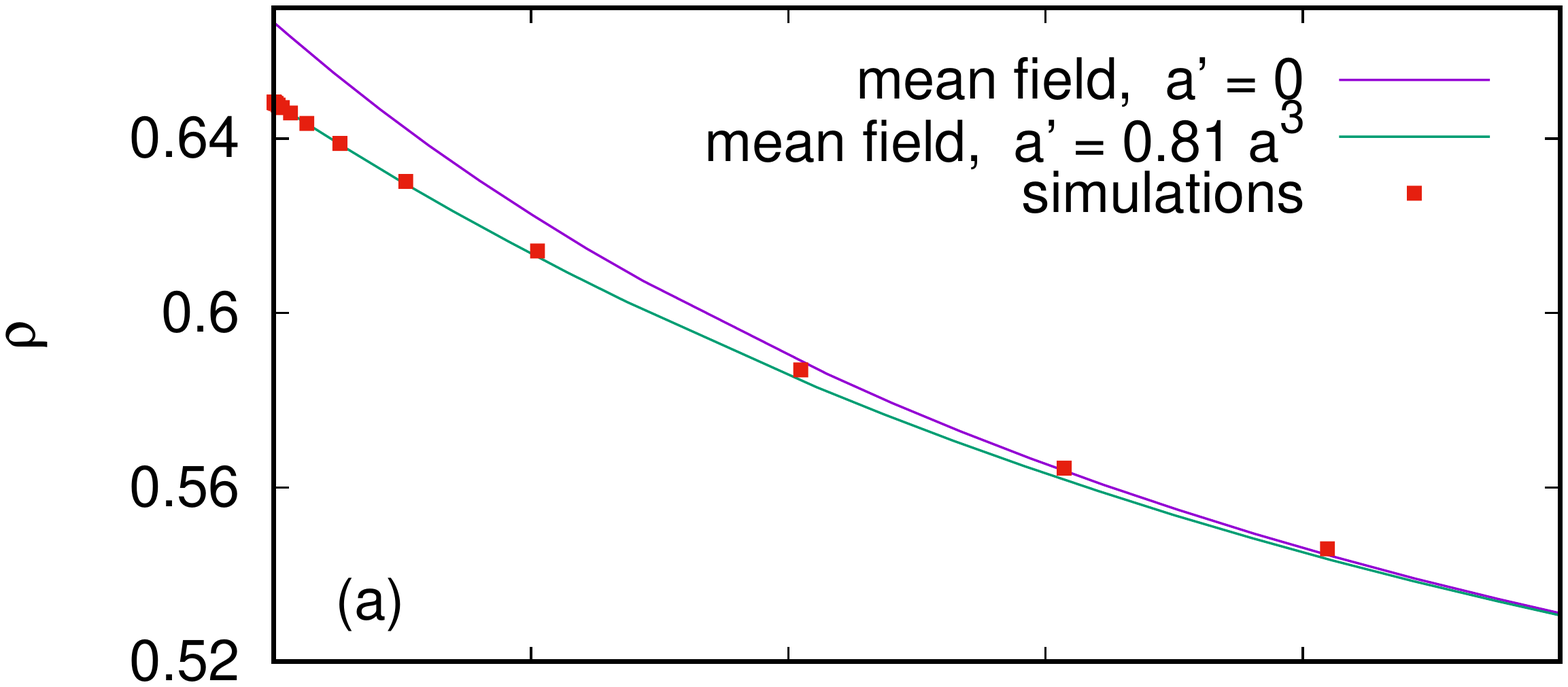}
   \vglue -3.2cm
   \includegraphics[width=0.50\textwidth]{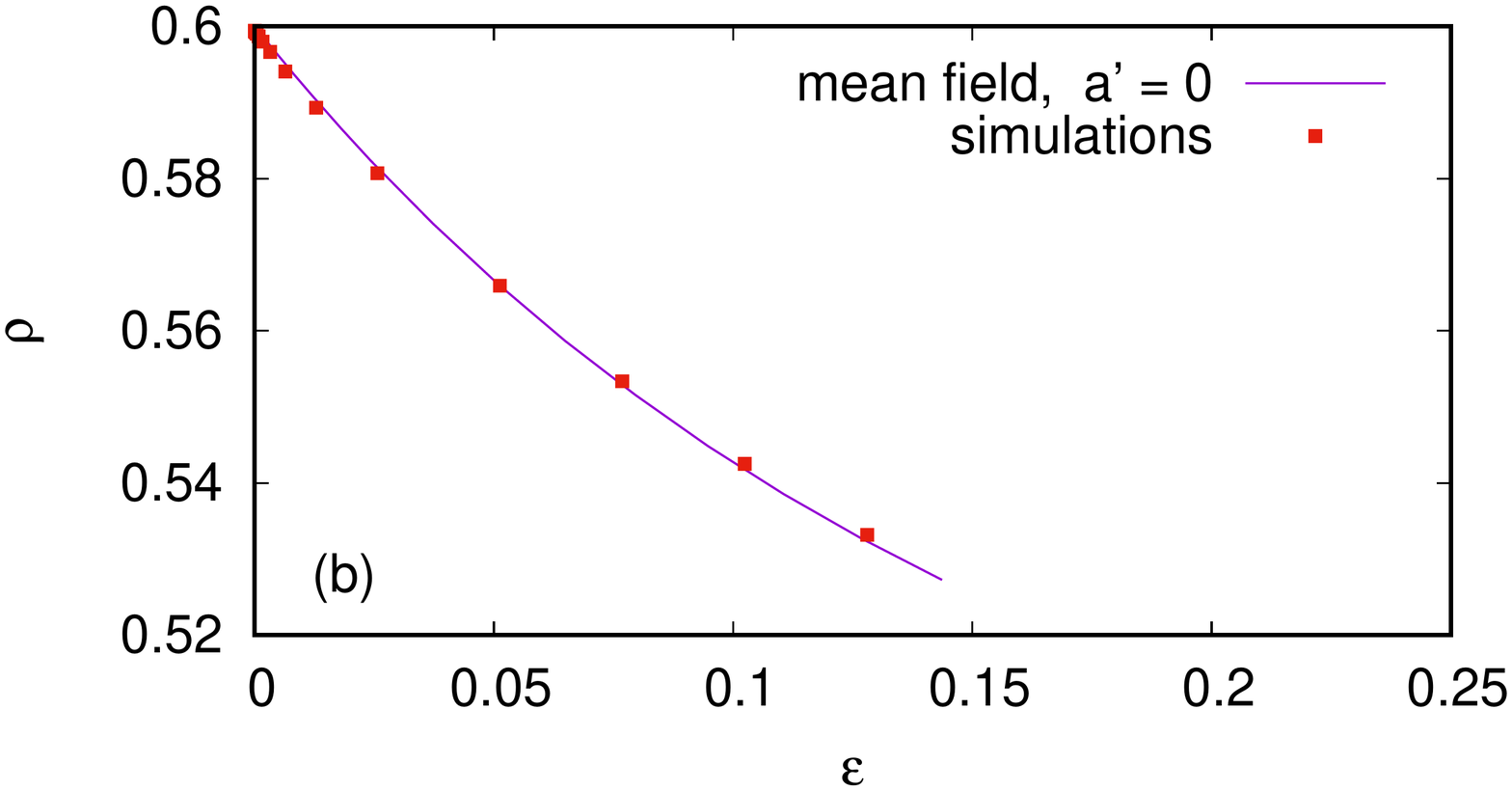}
\end{center}
   \vglue -0.9cm
	\caption{\small Stationary densities (average spins), measured between events.
	Statistical errors are much smaller than the data points. For $d=2$ (panel(a))
	we show also two mean field predictions, one (the upper smooth curve) for $a=0$,
	the other for $a' = 0.81 a^3$. For $d=3$ (panel (b)) we only show the mean field curve for 
	$a'=0$, as this gives already a nearly perfect fit.}
        \label{density.fig}
\end{figure}

\begin{figure}
\begin{center}
   \vglue -1.4cm
   \includegraphics[width=0.50\textwidth]{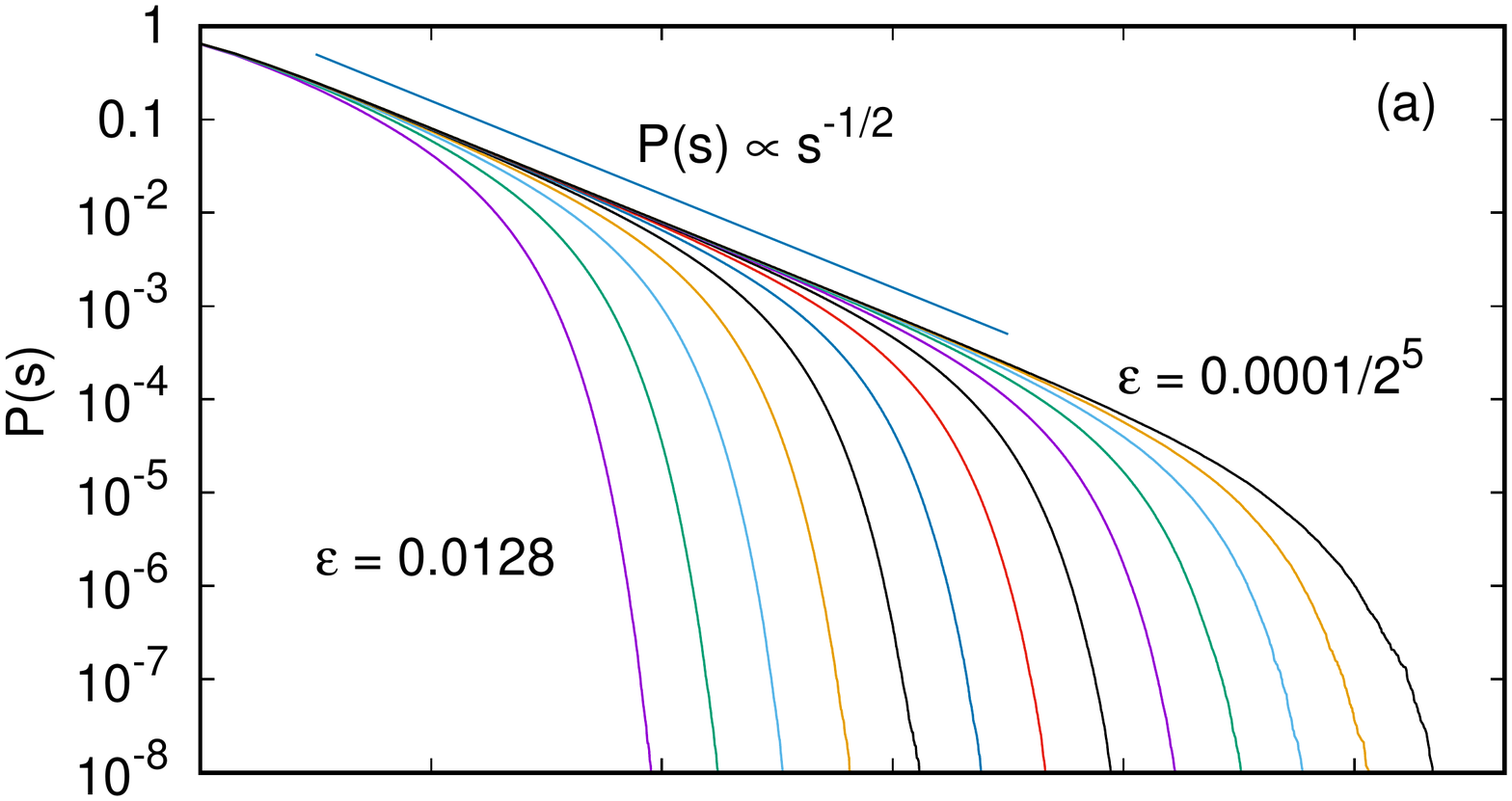}
   \vglue -2.4cm
   \includegraphics[width=0.50\textwidth]{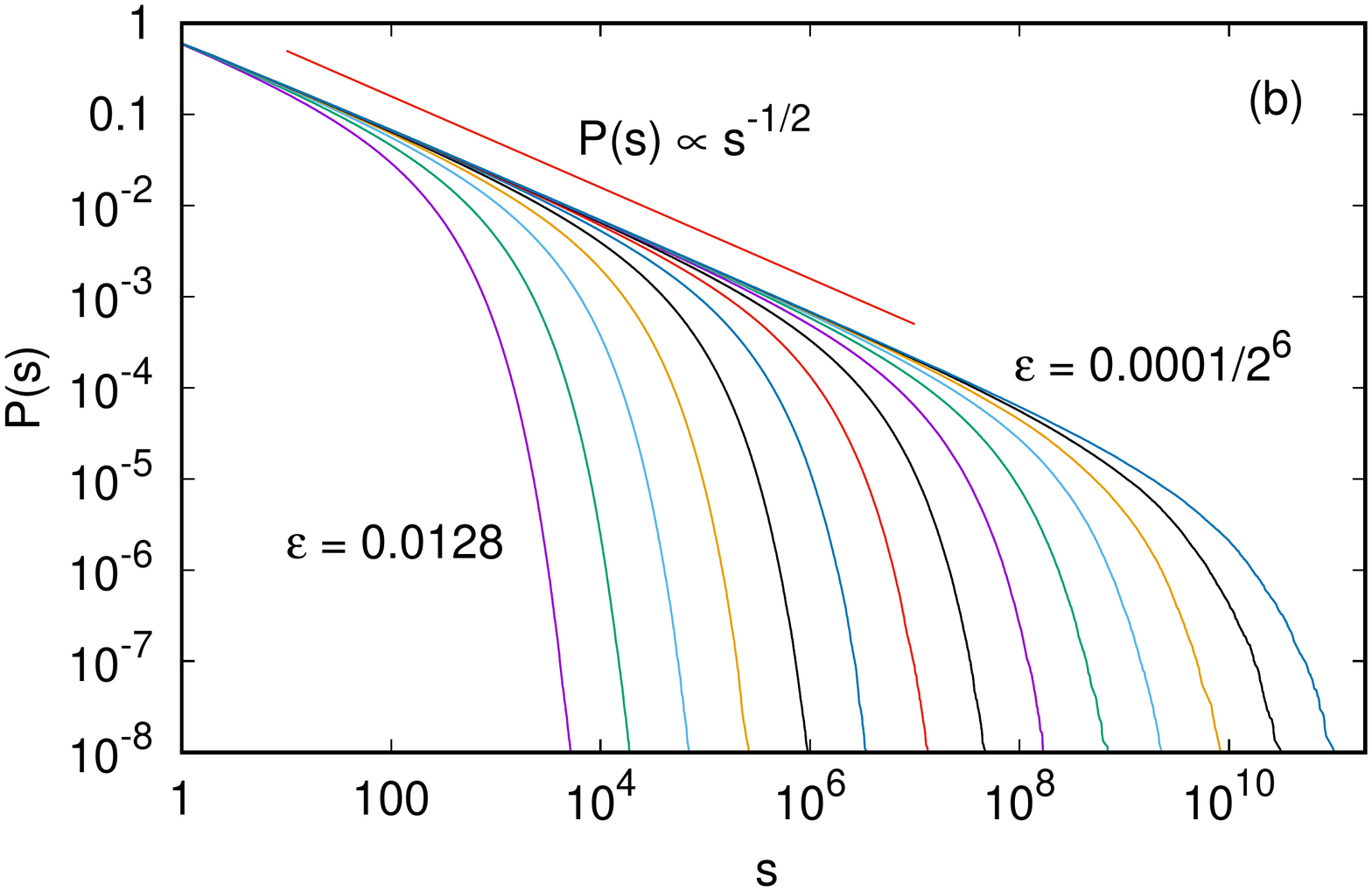}
\end{center}
   \vglue -0.9cm
	\caption{\small Log-log plots of the cumulative distributions of the number of 
	topplings per event. 
	Here and in the following plots, the curves are for $\epsilon = \epsilon_{\rm min},
	2\epsilon_{\rm min},4\epsilon_{\rm min},\ldots \epsilon_{\rm max}$.
	Statistical errors are significant only in the far right tails, as indicated by
	the (very small) fluctuations. The similarity between the two sets of curves 
	reflects the fact that both are essentially those for the mean field model.}
        \label{hist.fig}
\end{figure}

\begin{figure}
\begin{center}
   \vglue -1.2cm
   \includegraphics[width=0.50\textwidth]{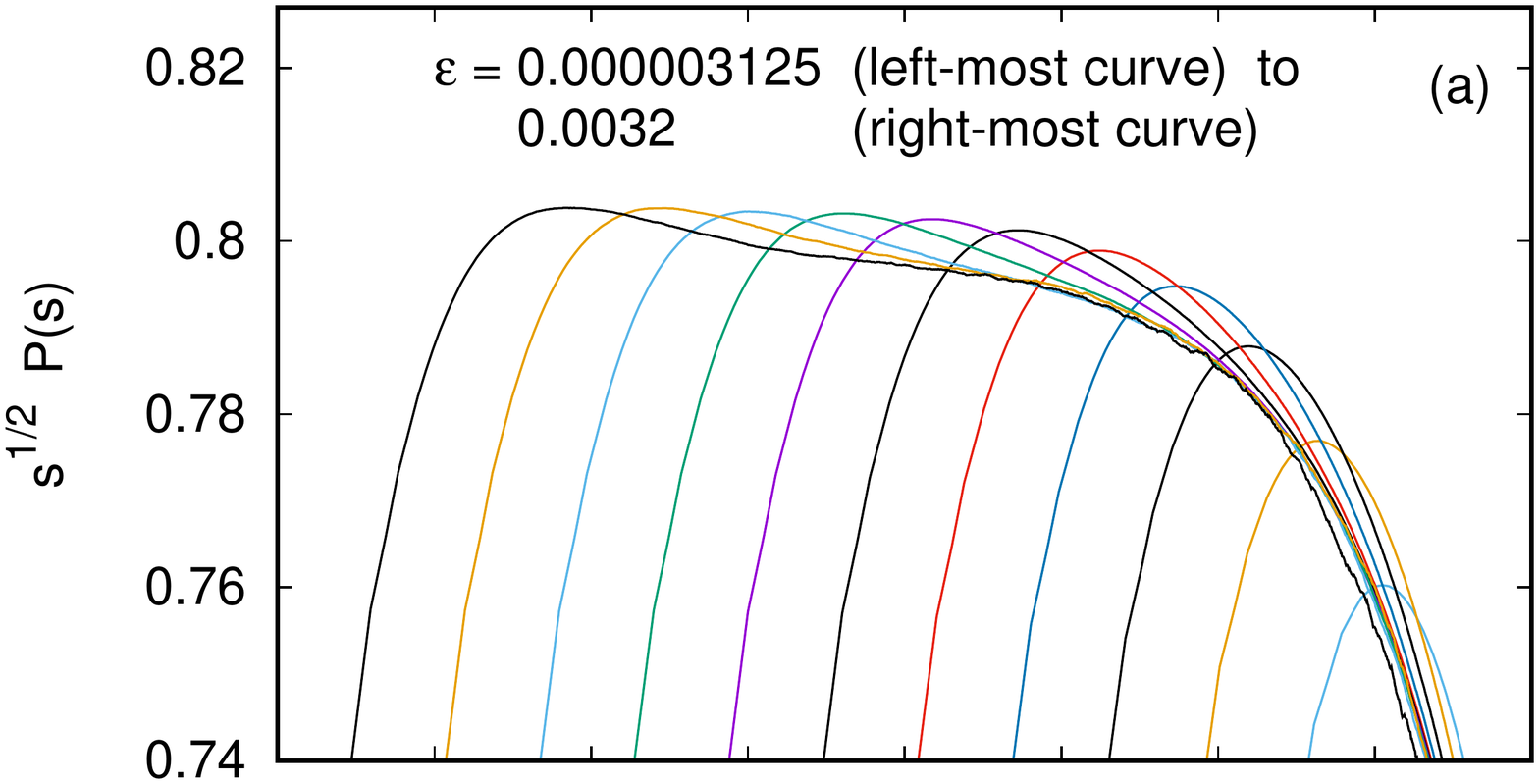}
   \vglue -2.4cm
   \includegraphics[width=0.50\textwidth]{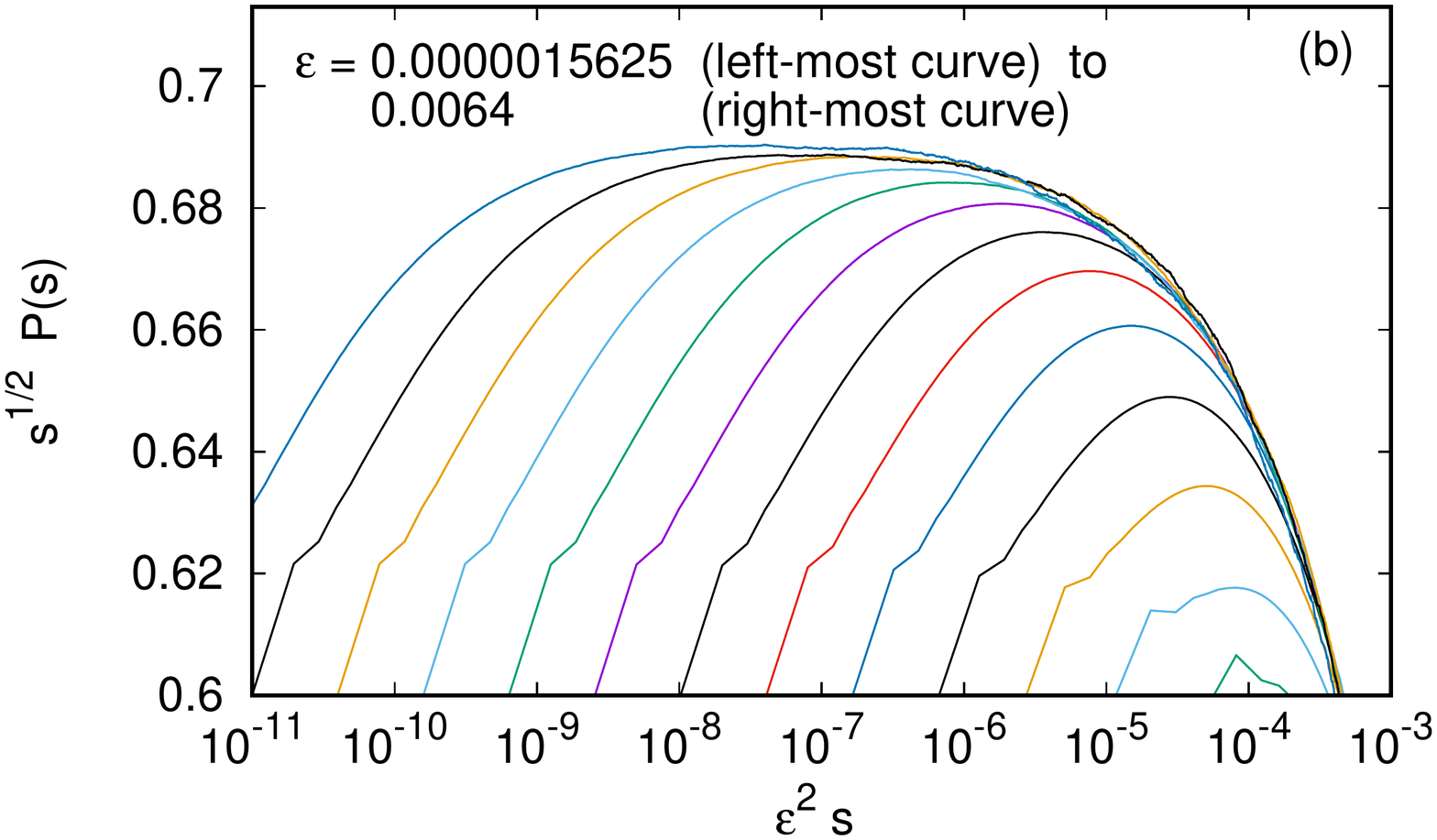}
\end{center}
   \vglue -0.9cm
	\caption{\small Same data as in Fig.~\Ref{hist.fig}, but plotted as $\sqrt{s}P(s)$
	against $\epsilon^2 s$. According to Eq.~(\Ref{P}), these curves should tend towards
	straight horizontal curves for $\epsilon\to 0$. To see more clearly whether this
	is true or not, we show the data on a strongly blown up (non-logarithmic) y-scale.}
        \label{hist-a.fig}
\end{figure}

\begin{figure}
\begin{center}
   \vglue -1.7cm
   \includegraphics[width=0.50\textwidth]{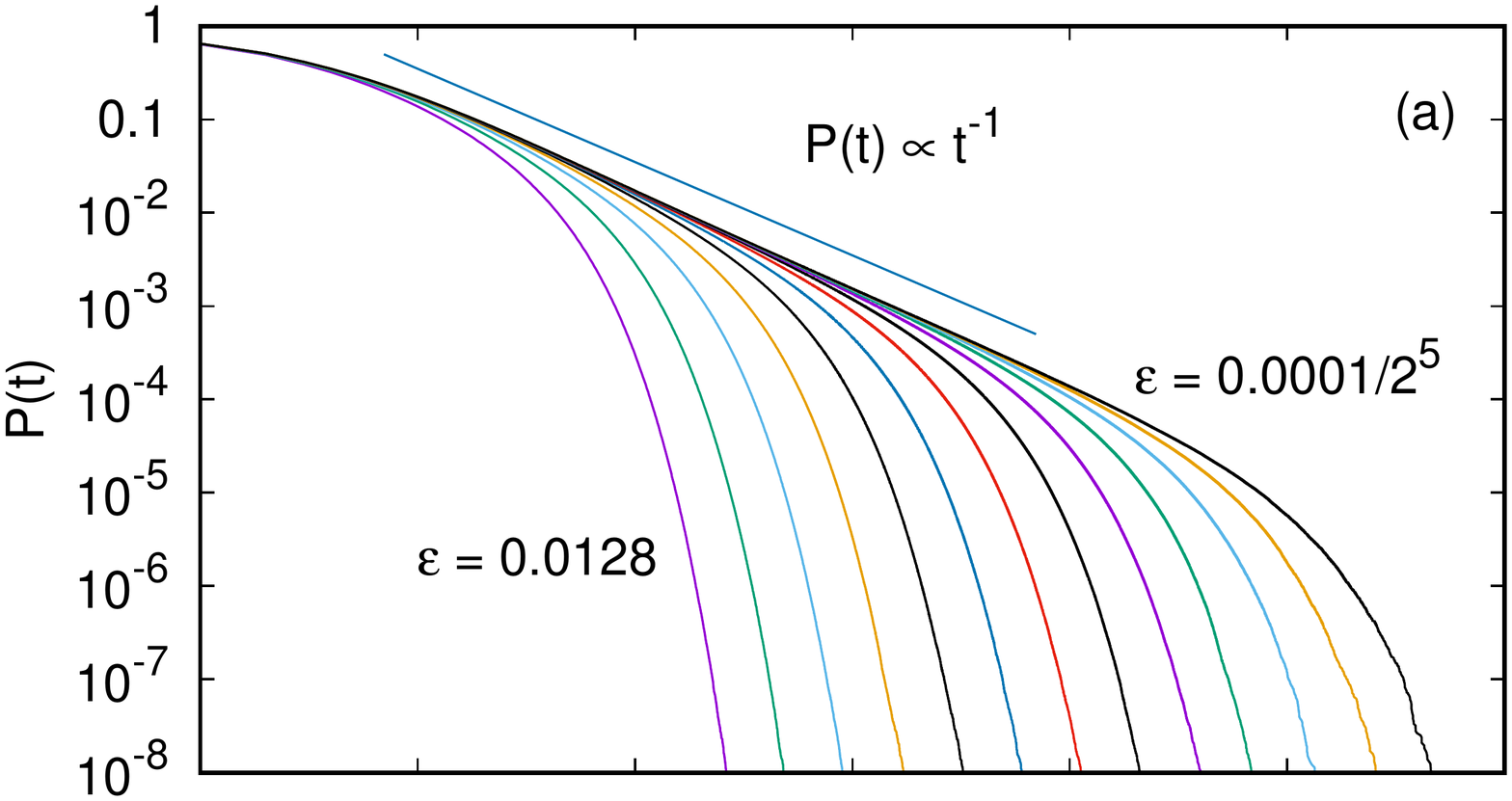}
   \vglue -2.4cm
   \includegraphics[width=0.50\textwidth]{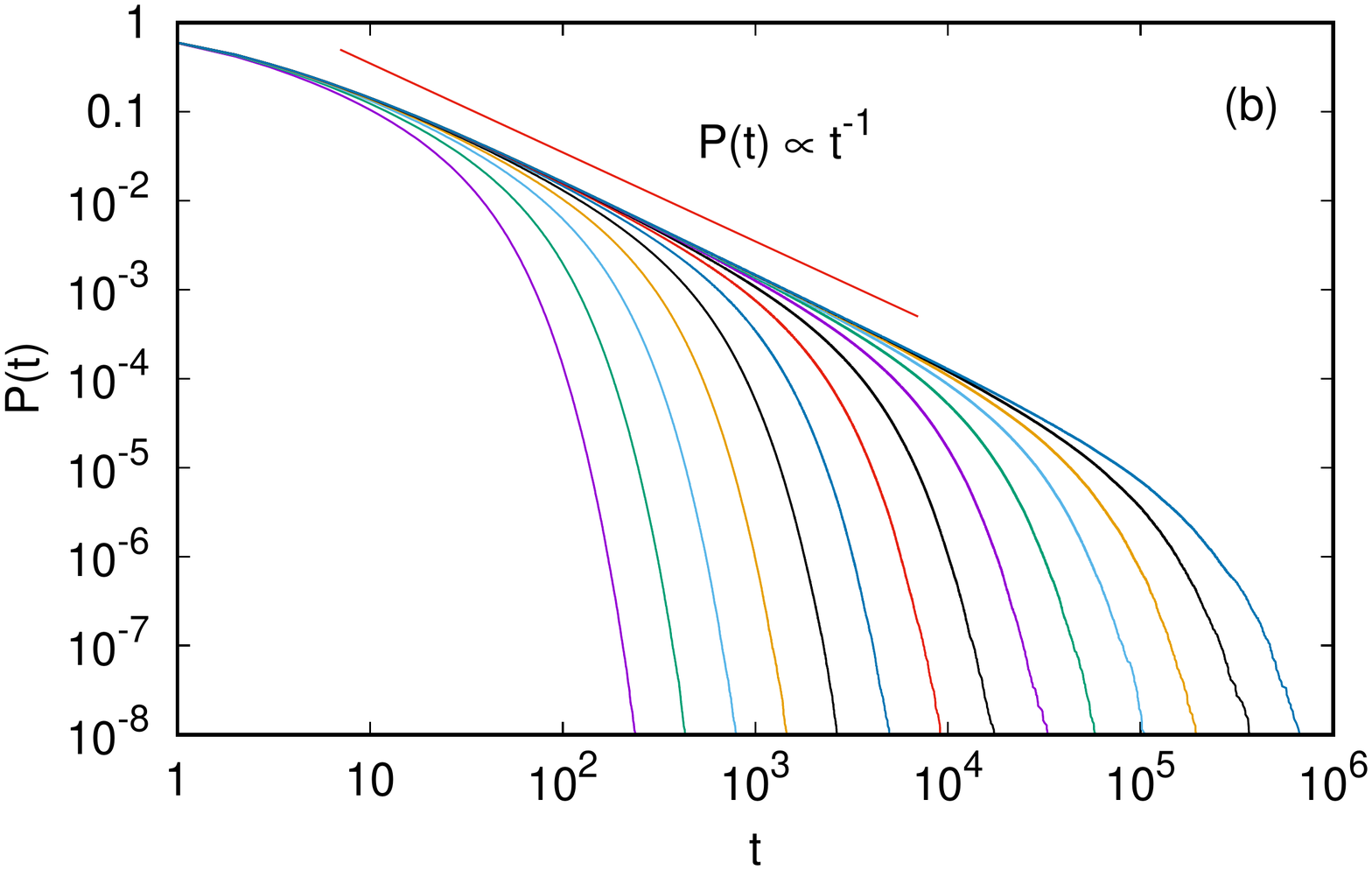}
\end{center}
   \vglue -0.9cm
        \caption{\small Log-log plots of the cumulative distributions of the durations $t$
	of avalanches. Again, statistical errors are significant only in the far right 
	tails, and the similarity between the two sets of curves reflects again the fact 
	that both are essentially those for the mean field model.}
        \label{thist.fig}
\end{figure}

\begin{figure}
\begin{center}
   \vglue -1.2cm
   \includegraphics[width=0.50\textwidth]{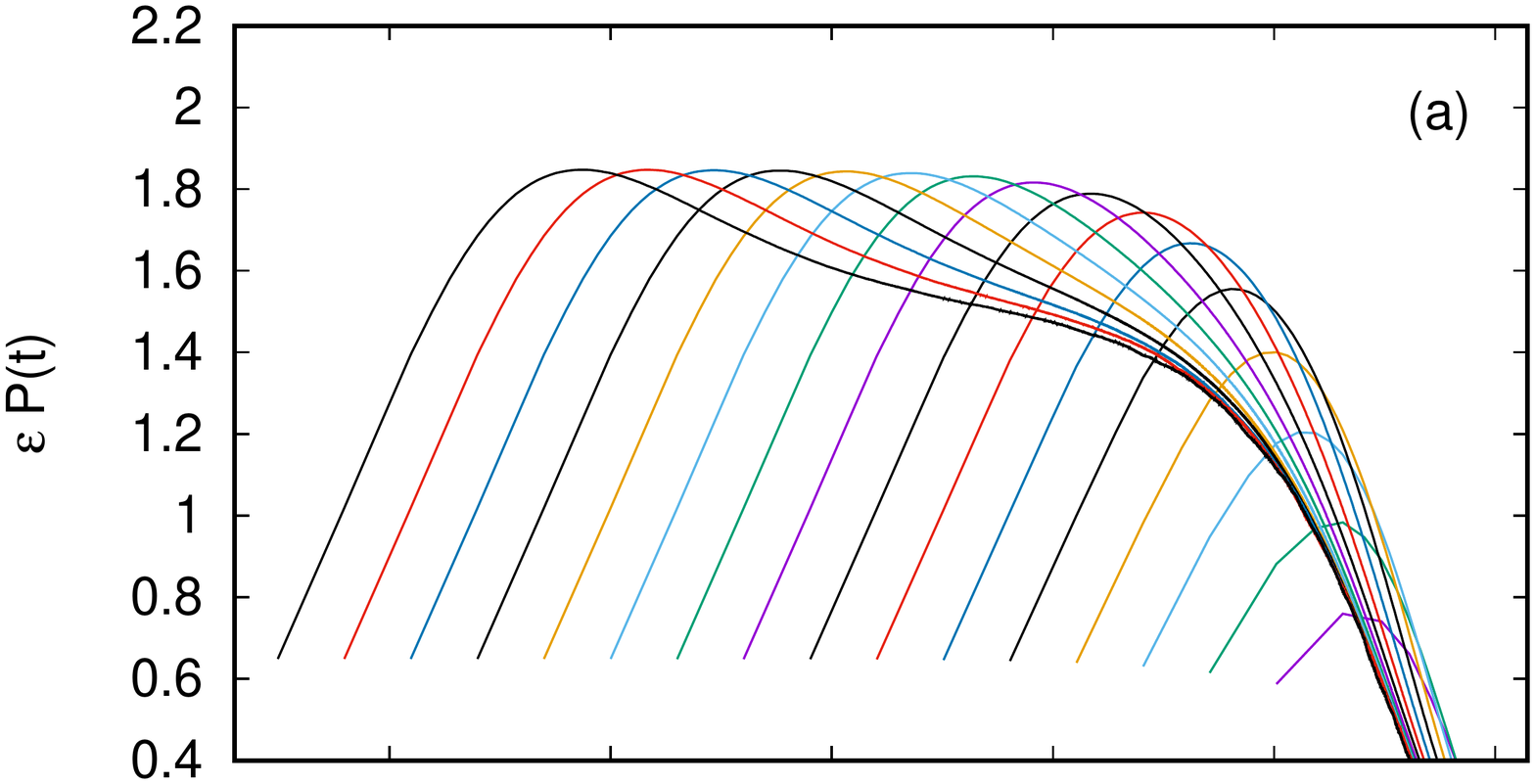}
   \vglue -2.6cm
   \includegraphics[width=0.50\textwidth]{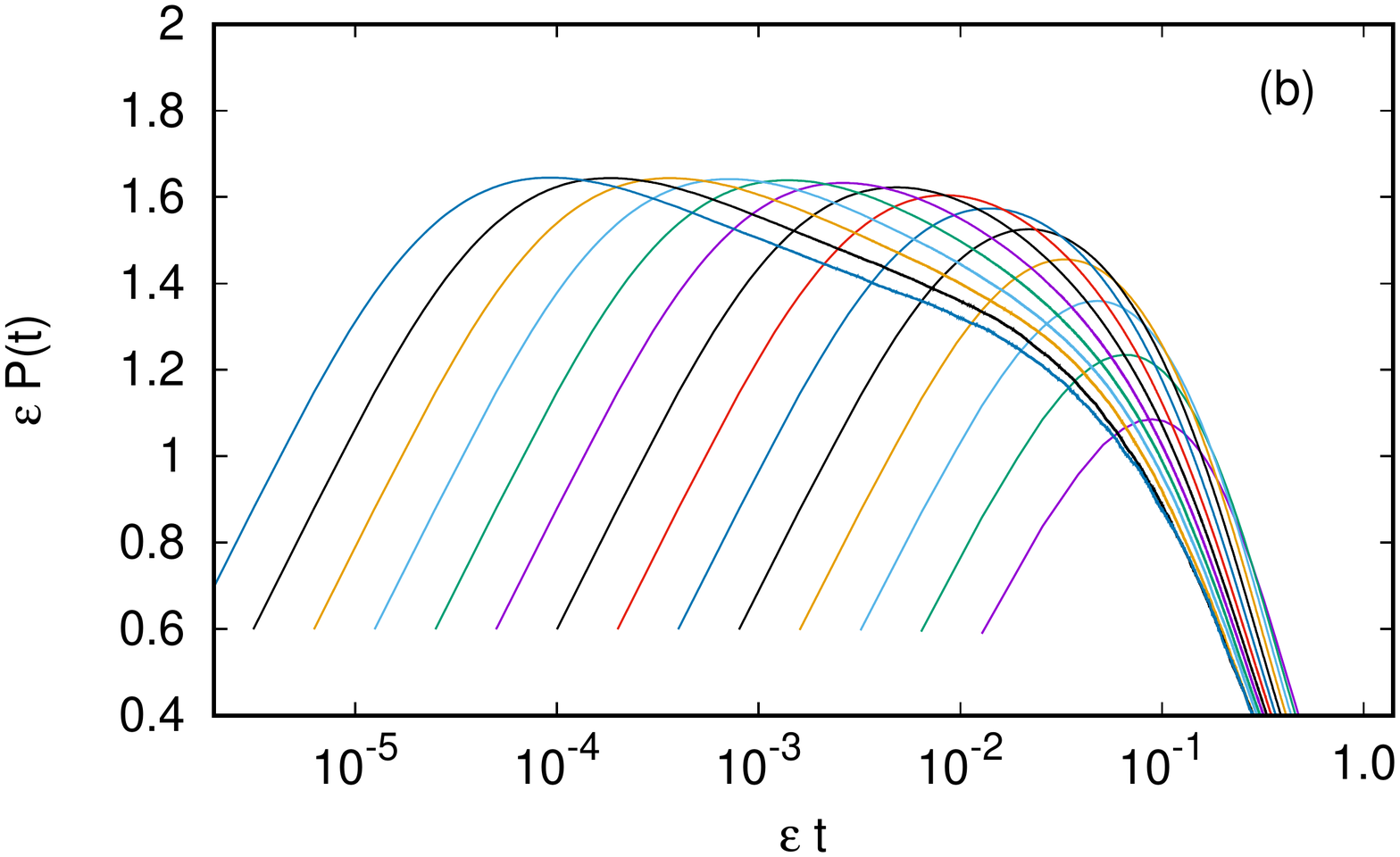}
\end{center}
   \vglue -0.9cm
	\caption{\small Same data as in Fig.~\Ref{thist.fig}, but plotted as $t Q(t)$
        against $\epsilon t$. According to Eq.~(\Ref{Q}), these curves should tend towards
        straight horizontal curves for $\epsilon\to 0$. To see more clearly whether this
        is true or not, we show the data on a strongly blown up (non-logarithmic) y-scale.}
        \label{thist-a.fig}
\end{figure}

Finally, we show in Figs.~5 to 8 the (cumulative) distributions $P(s)$ of the number $s$ 
of topplings per event and $Q(t)$ of the time durations $t$. We could make detailed 
comparison with the mean field model as in \cite{Broker}, but we prefer to just show 
that the scaling predictions 
\be
   P(s) = \frac{1}{\sqrt{s}}\Psi(\epsilon^2 s)    \label{P}
\ee
and 
\be
   Q(t) = \frac{1}{t} \Phi(\epsilon t)      \label{Q}
\ee
are excellently fulfilled in the asymptotic region $\epsilon \to 0$. For both we show
(in Figs.~5 and 7) first the unmodified distributions in their entire ranges, but 
since this is not really significant we then plot (in Figs.~6 and 8) blow-ups of the 
scaling parts for $\sqrt{s} P(s)$ versus $\epsilon^2 n$ and for $tQ(t)$ versus $\epsilon t$.
In all four plots (for $d=2$ and $d=3$, and for $P$ and $Q$) we see clear deviations 
from scaling (i.e., none of the curves are horizontal in the central regions), but in 
all four cases these seem clearly to disappear for $\epsilon \to 0$. 

\begin{figure}[htp]
\begin{center}
   \vglue -0.1cm
   \includegraphics[width=0.43\textwidth]{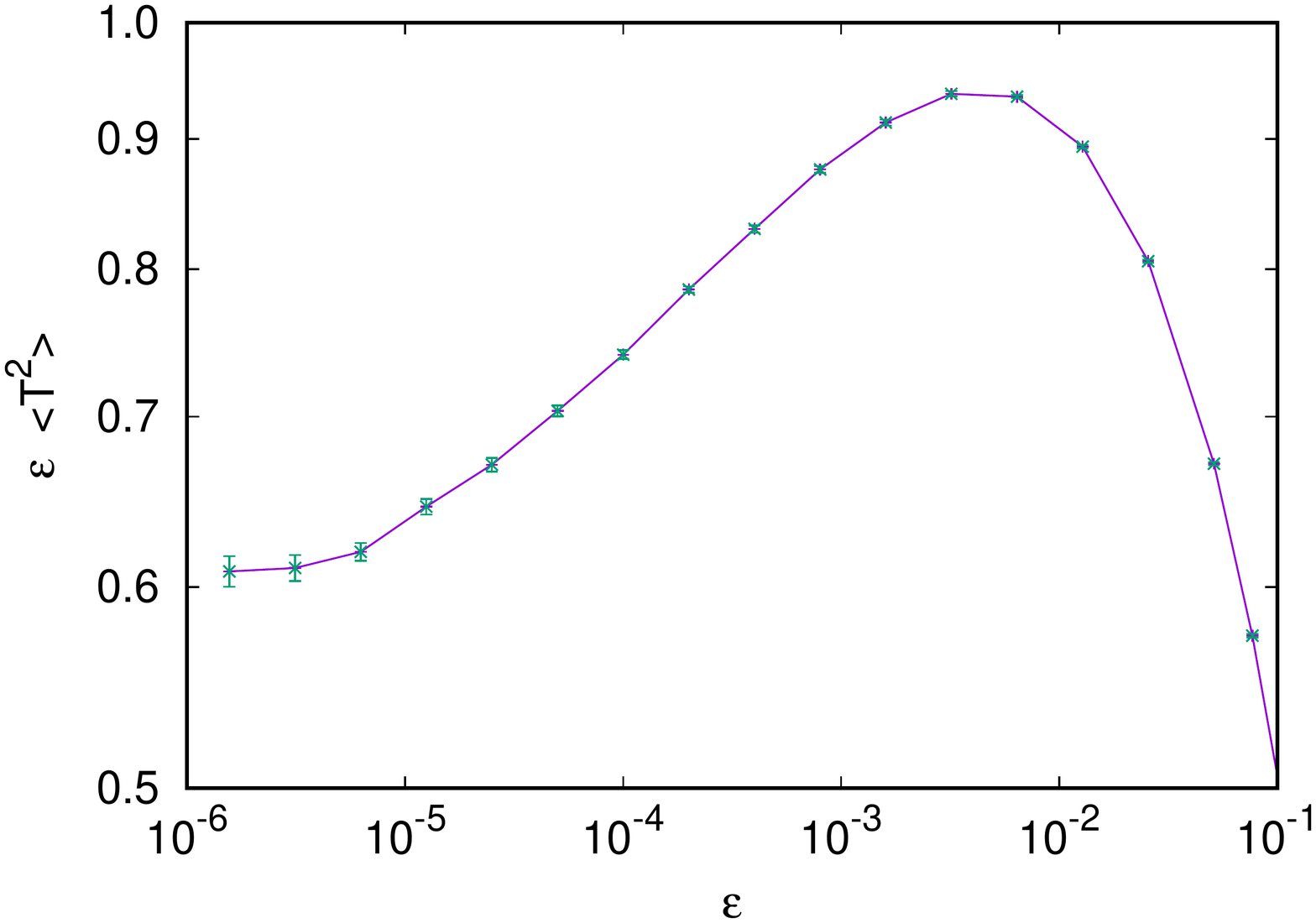}
\end{center}
   \vglue -0.9cm
	\caption{\small Log-linear plot of the average squared avalanche durations in $d=3$
	versus $\epsilon$. For increased significance we show $\epsilon\langle t^2\rangle$
	instead of $\langle t^2\rangle$ itself. We see large corrections to scaling, but 
	for $\epsilon\to 0$ the curve becomes horizontal as predicted by Eq.~(\Ref{Q}).}
        \label{3d-T_averages-a.fig}
\end{figure}

In Fig.~8b we also
see quite a substantial violation of scaling for large values of $t$ (curves do not 
collapse there), but a closer inspection shows that this also disappears for 
$\epsilon \to 0$. This is also seen more clearly from Fig.~9, where we show 
$\langle t^2\rangle$ for $d=3$. Eq.(\Ref{Q}) implies that $\langle t^2\rangle \propto 
1/\epsilon$ or 
\be
   \epsilon \langle t^2\rangle = const,
\ee
but we see that this only becomes true for extremely small values of $\epsilon$. 
Basically the same happens in $d=2$, although the scaling violations there are much 
smaller (data not shown). It is precisely the scaling violations seen in Figs.~6,8, and 9
which were at the basis of claims in \cite{Broker} that mean field scaling is not 
exact, because at that time we were unable to simulate at as small values of 
$\epsilon$ as in the present paper.

In summary, we have shown that the critical scaling in the model of \cite{Broker} is indeed
precisely that of its mean field version, both for $d=2$ and $d=3$. To our knowledge this is 
the first and only model in low dimensions and with short-range interactions where this was
ever observed. Indeed, mean field scaling usually does not occur in such models, if they 
show detailed balance, i.e. if they describe equilibrium phenomena, since feedback loops 
there tend to have a definite sign. Take, e.g., the Ising model. Loops contribute obviously
with a positive sign in the ferromagnetic Ising model, but also in the antiferromagnetic
one on bipartite lattices. The same is true for other models like self avoiding walks, the 
Heisenberg model, and percolation. We are not aware of a proof that the same holds for all
equilibrium models, thus a search for mean field behavior in equilibrium critical phenomena
might be worth while. In the present (non-equilibrium) model, these arguments do not apply 
because multiple topplings lead to `spin' changes of either signs, and thus to cancellations
in the contribution of loops. As in \cite{Broker}, we propose to call this phenomenon
``confusion". It can be viewed as the first known instance where the basic concept of the random 
phase approximation \cite{Bohm,GellMann} becomes exact, and even that only at a critical point.


\begin{thebibliography}{99}
\bibitem{footnote1} This model had actually been mentioned first as model \#5 in the 
	appendix of \cite{Grass_nr5},
	because its depth-first (recursive) simulation leads to an algorithm very close to 
	percolation, to self-avoiding walks, and to the Wolff algorithm for the Ising model.
\bibitem{footnote2} Since the model is Abelian in the sense of \cite{Dhar}, the order in 
	which these topplings are performed is irrelevant.
\end{thebibliography}
\end{document}